\documentclass[aps,prb,reprint,unsortedaddress,superscriptaddress]{revtex4-1}

\usepackage{amsmath,amssymb}

\usepackage[dvips]{graphicx}

\usepackage{bm}
\usepackage{color}

\begin{document}

\title{Spherical topological insulator}

\begin{abstract}
The electronic spectrum on the spherical surface of a topological insulator
reflects an {\it active} property of the helical surface state 
that stems from a constraint on its spin on a curved surface.
The induced spin connection can be interpreted
as an effective vector potential 
associated with a fictitious magnetic monopole
induced at the center of the sphere.
The strength of the induced magnetic monopole is found to be
$g=\pm 2\pi$,
being the smallest finite (absolute) value compatible with the Dirac quantization condition.
We have established an explicit correspondence between the bulk Hamiltonian 
and the effective Dirac operator on the curved spherical surface.
An explicit construction of the surface spinor wave functions implies
a rich spin texture possibly realized on the surface of topological insulator nanoparticles. 
The electronic spectrum inferred by the obtained effective surface Dirac theory,
confirmed also by the bulk tight-binding calculation,
suggests a specific photo absorption/emission spectrum of such nanoparticles.
\end{abstract}

\author{Ken-Ichiro Imura}
\affiliation{Department of Quantum Matter, AdSM, Hiroshima University, Higashi-Hiroshima 739-8530, Japan}
\author{Yukinori Yoshimura}
\affiliation{Department of Quantum Matter, AdSM, Hiroshima University, Higashi-Hiroshima 739-8530, Japan}
\author{Yositake Takane}
\affiliation{Department of Quantum Matter, AdSM, Hiroshima University, Higashi-Hiroshima 739-8530, Japan}
\author{Takahiro Fukui}
\affiliation{Department of Physics, Ibaraki University, Mito 310-8512, Japan}

\date{\today}

\maketitle

\section{Introduction}

It was only several years ago 
that the idea of topological insulator
has been proposed as a possible candidate for the new state of matter
in the field of condensed-matter.
\cite{Moore_review}
The original theoretical idea has already been extended in various aspects,
made applicable to a broader range of phenomena, including superconductivity
and superfluidity.
\cite{QiZhang,YTanaka}
The related research areas are now reclassified and recognized as that of 
the topological quantum phenomena.
Naturally, the outbreak of this new research field owes much to 
a rapid success of experimental studies
that have demonstrated that the new theoretical idea has much reality.
\cite{HasanKane}

The existence of a single gapless Dirac cone in its surface spectrum
is a hallmark of strong topological insulators.
Here, we focus on a specific property of this robust and protected surface state
on a curved surface, 
\cite{DHLee}
the ''spin-to-surface locking''.
It is indeed specific to the topological insulator surface state and distinguish it
from other realizations of gapless Dirac cones in condensed matter
such as in graphene \cite{Geim, TAndo} and related carbon materials.
The role of spin-to-surface locking may be most accentuated in the (pseudo-cylindrical)
wire-shaped geometry
in which an anomalous Aharonov-Bohm type of oscillation has been reported.
\cite{AB_exp}
Motivated by the reality of such transport measurements 
which may allow for a direct observation of the spin Berry phase,
theorists have extensively studied the role of this phenomenon
in the transport characteristics of the surface state.
\cite{Ashvin_PRL, Mirlin_PRL, JMoore_PRL, k1, k2}

A remarkable consequence of the spin-to-surface locking in the
cylindrical geometry is the {\it half-integral} quantization of the
{\it orbital} angular momentum.
Clearly, such half-integral quantization 
leads to appearance of a finite-size energy gap 
in the hitherto gapless surface electronic spectrum. 
Interestingly, 
introduction of 
a physical magnetic flux of half of a unit flux quantum
through (piercing) the cylinder
compensates the Berry phase associated with the spin-to-surface locking,
and closes the gap.
The same mechanism applies to the classification 
of gapless electronic states
bound to a crystal dislocation line 
penetrating an otherwise 
surfaceless sample of 
a three-dimensional topological insulator.\cite{Ran}
A more systematic consideration
\cite{TeoKane}
on such gapless electronic states
associated with a topological defect in a topological mother system
has been developed from the viewpoint of 
classifying topological insulators and superconductors
in a unified way 
solely from their symmetry class.
\cite{Schnyder_PRB, Kitaev_AIP, Schnyder_AIP, Ryu_NJP}

The specificity of the cylindrical surface is that it is {\it flat}
in the sense that it has everywhere a vanishing Gaussian curvature.
On the surface of a topological insulator of more generic shape or
geometry yielding a finite curvature,
the effect of spin-to-surface locking mentioned earlier
will be modified by that of a finite curvature.
A spherical surface of topological insulator
\cite{Guinea_2011}
is a prototypical example in which
such an interplay is expected.
We show in this paper that the two effects are both expressed
in terms of a Berry phase, 
but of contrasting nature (see Table \ref{table_Berry}).
The two types of Berry phase both contribute
to the formation of a finite-size energy gap.
The resulting surface electronic spectrum on the sphere is shown 
to have a substantial compatibility 
with the result of tight-binding calculations
performed for a cubic system
(for tight-binding calculation involving the bulk,
cubic implementation is much straightforward).
A related but different scenario on
the fate of such a (planar) gapless Dirac cone embedded 
on the curved spherical surface
has been proposed
in the study of the electronic states in
fullerene.
\cite{Guinea_PRL, Guinea_NPB, Abrikosov_arxiv, Abrikosov, EPJB_fullerene}

In addition to the spectrum, the structure of the surface spinor wave function
is another highlight of the paper.
On the curved spherical surface of a topological insulator
the strong spin-orbit coupling in the bulk,
combined with the twisting of the phase shift
due to the two types of Berry curvature,
leads to a non-trivial {\it spin texture}.
By explicitly constructing the surface spinor wave function
we reveal such a rich spin texture
possibly realized on the surface of topological insulator
nano-particles.

The paper is organized as follows.
In Sec. II, the effective surface Dirac theory is derived
from the gapped bulk Hamiltonian,
in which
two types of Berry phase appear.
The nature of
these two types of Berry phase
is discussed and contrasted in Sec. III.
The solution of the effective surface Dirac equation
is given explicitly in Sec. IV.
The surface wave function is shown to be expressed in terms of the
Jacobi's polynomials.
The obtained discrete energy spectrum is compared with the result of
(bulk) tight-binding calculation in Sec. V.
This leads us to our conclusions.
Some details of the formulation are left to the appendices.

\begin{table*}[htdp]
\caption{Characterization of the two types of Berry phase manifesting on the surface of a spherical
topological insulator}
\begin{center}
\begin{tabular}{lll}
\hline\hline
type & (A) & (B) \\
\hline
(geometrical) origin & curvature in the polar ($\theta$-) direction; effect of & effect of rolling the surface in the azimuthal 
\\
& closing the surface at the north and south poles &  ($\phi$-) direction \\
\hline
appearance & in the covariant derivatives, or & $-i\partial_\phi \rightarrow -i{\partial}_\phi +1/2$ \\
(where, how) & $\partial_\theta \rightarrow \partial_\theta + {1\over 2} \cot (\theta/2)$ & \\
\hline
shifting the spectrum? & yes & yes \\
\hline
relation to spin-to-surface & breaks the locking & expression of the (tendency to)\\
locking & & spin-to-surface locking \\
\hline
sensitivity to the choice & no & yes \\
of basis & & \\
\hline
other examples? & fullerene (buckyball) & cylindrical TI\\
\hline\hline
\end{tabular}
\end{center}
\label{table_Berry}
\end{table*}

\section{Derivation of the surface effective Hamiltonian}
Let us first derive
an effective ``Dirac operator''
on the spherical surface,
starting with a bulk Hamiltonian.
Our starting point is the following gapped bulk effective Hamiltonian
\cite{Liu_nphys, Liu_PRB}
in the continuum limit:
\begin{equation}
H_{\rm bulk}=\epsilon (\bm p) \bm 1
+ m (\bm p) \tau_z + A\tau_x (p_x \sigma_x + p_y \sigma_y + p_z \sigma_z),
\label{H_bulk}
\end{equation}
describing a three-dimensional (3D) $\mathbb Z_2$ topological insulator, 
where
\begin{equation}
m(\bm p) = m_0 + m_2 (p_x^2 + p_y^2 + p_z^2),
\label{mass_conti}
\end{equation}
is a (generalized) mass term containing both
the constant and quadratic (Wilson) terms.
For simplicity,
we have chosen the Wilson term to be isotropic.
The two types of Pauli matrices
$\bm\sigma =(\sigma_x, \sigma_y, \sigma_z)$ and
$\bm\tau =(\tau_x, \tau_y, \tau_z)$
represent, respectively,
the real and orbital spin degrees of freedom,
and $\bm 1$ is the $4\times 4$ identity matrix.
The Hamiltonian (\ref{H_bulk}) is time-reversal invariant,
i.e., invariant under the operation of
$\Theta = i\sigma_y K$,
where $K$ represents complex conjugation.
The two types of Pauli matrices represent independent degrees of freedom
acting on spinors living in a different space.
To make this point explicit one may express Eq. (\ref{H_bulk})
in the following $4\times 4$ matricial form,
\begin{equation}
H_{\rm bulk}=\epsilon (\bm p) \bm 1+
\left[
\begin{array}{cccc}
m (\bm p) & A p_z  & 0 & A p_- \\
A p_z & - m (\bm p) & A p_- & 0 \\
0 & A p_+ & m (\bm p) & - A p_z \\ 
A p_+ & 0 & - A p_z & - m (\bm p)   
\end{array}
\label{matH}
\right],
\end{equation}
where $p_\pm = p_x \pm i p_y$.

In the following demonstration,
we choose $\epsilon (\bm p)$ to be null
so that the spectrum be symmetric with respect to $E=0$
(particle-hole symmteric).
The vanishing of the $\epsilon (\bm p) \bm 1$ term
upgrades the symmetry of the model
from class AII to DIII (see Appendix A for details),
but leaves unchanged the distinction 
between topologically trivial ($m_0/m_2>0$) and 
non-trivial ($m_0/m_2>0$) phases.
Note that
the minimal model we consider contains only three control parameters,
$m_0$, $m_2$ and $A$.
We also consider the spherical geometry, assuming that
the topological insulator 
described by Eqs. (\ref{H_bulk}) and (\ref{mass_conti}) 
occupies the interior of a sphere of radius $R$.
We introduce standard 3D spherical coordinates: $(r, \theta, \phi)$ 
related to the cartesian coordinates as
\begin{eqnarray}
x&=&r\sin\theta \cos\phi,
\nonumber \\
y&=&r\sin\theta \sin\phi,
\nonumber \\
z&=&r\cos\theta.
\end{eqnarray}
The momentum components, $(p_x, p_y, p_z)$ 
in Eq. (\ref{matH}) 
expressed in the cartesian coordinates
are rewritten in terms of the derivatives 
with respect to the spherical coordinates, $(r, \theta, \phi)$,
by following the standard procedure.
We also introduce unit vectors,
$\hat{\bm r}$, $\hat{\bm\theta}$ and $\hat{\bm\phi}$,
pointed, respectively, in the direction of
the increase of $(r, \theta, \phi)$.
The momentum operator 
$\bm p=p_x \hat{\bm x}+p_y \hat{\bm y}+p_z \hat{\bm z}$
can be reprojected onto the directions of such unit vectors
in the spherical coordinates as
$\bm p=p_r \hat{\bm r}+p_\theta \hat{\bm\theta}+p_\phi \hat{\bm\phi}$,
where
$p_r = -i(\partial_r + 1/r)$, $p_\theta = -i\partial_\theta /r$ and
$p_\phi = -i\partial_\phi /(r\sin\theta)$.

To derive the surface effective Hamiltonian
in the spirit of $k\cdot p$-approximation,
we divide $H_{\rm bulk}$ into two parts:
$H_{\rm bulk} = H_\perp + H_\parallel$,
where
$H_\perp = H |_{p_\theta =0, p_\phi=0}$,
and first solve the radial eigenvlaue problem:
\begin{equation}
H_\perp |\psi \rangle = E_\perp |\psi \rangle,
\label{radial}
\end{equation}
instead of
$H_{\rm bulk} |\psi \rangle = E |\psi \rangle$.
Let us consider 
a possible form of
the surface solutions of Eq. (\ref{radial}).
$|\psi \rangle$ may take the following form:
\begin{equation}
|\psi \rangle =|\psi (r, \theta, \phi) \rangle = e^{\kappa (r-R)} |u (\theta, \phi) \rangle,
\label{damped}
\end{equation}
where
$\kappa^{-1}$ measures the penetration of the surface wave function into the bulk.
Here,
taking a linear combination of the solutions
of the form of Eq. (\ref{damped})
we construct a solution of
Eq. (\ref{radial})
which is compatible with the boundary condition:
\cite{Liu_nphys, Liu_PRB, Shen_NJP, k2}
\begin{equation}
|\psi (r=R)\rangle=\bm 0,
\label{bc}
\end{equation}
i.e., the condition that all the four components of the wave function
$\psi$ vanish on the surface of the sphere (at $r=R$).
As shown in Appendix B,
this can be matched by
superposing two damped solutions of the form of
Eq. (\ref{damped}).
Importantly,
the solutions of such a boundary value problem must satisfy
the zero-energy condition (\ref{zero}),
i.e., the (surface) Dirac point is at $E=0$ in our model.



The zero-energy condition (\ref{zero}) 
helps simplify
the solution of the radial eigenvalue problem (\ref{radial}).
Since $E_\perp =0$, solving Eq. (\ref{radial}) is equivalent to finding $\psi$
that satisfies,\cite{BHZ_JPSJ, BHZ_zigzag}
\begin{equation}
\tau_z H_\perp |\psi \rangle = 0,
\label{tau_z}
\end{equation}
where for $|\psi \rangle$ taking the form of Eq. (\ref{damped}),
\begin{equation}
\tau_z H_\perp |\psi \rangle = [m_\perp [\kappa] + \tau_y A\kappa\hat{\bm r}\cdot\bm\sigma]|\psi \rangle,
\label{radial_2}
\end{equation}
where $m_\perp [\kappa] = m_0 - m_2 \kappa^2$.
This implies
that the orbital part of the eigenspinor $u (\theta, \phi)$ can be chosen as
an eigenstate of $\tau_y$,
\begin{equation}
\tau_y |\tau_y \pm\rangle = \pm |\tau_y \pm\rangle.
\end{equation}
To fix the notation,
let us express the explicit vectorial representation of 
$|\tau_y \pm\rangle$ as
\begin{equation}
|\tau_y +\rangle = {1\over \sqrt{2}}
\left[
\begin{array}{l}
1\\ i
\end{array}
\right],\ \ \
|\tau_y -\rangle = {1\over \sqrt{2}}
\left[
\begin{array}{l}
1\\ -i
\end{array}
\right].
\end{equation}
The real spin part of Eq. (\ref{radial_2})
can be also diagonalized 
by pointing the eigenstates of $\sigma_z$
in the direction of $\hat{\bm r}$, i.e., by
\begin{eqnarray}
|\hat{\bm r} +\rangle &=& {1\over \sqrt{2}}
\left[
\begin{array}{r}
e^{-i\phi /2}\cos{\theta \over 2} \\
e^{i\phi /2}\sin{\theta \over 2}
\end{array}
\right],
\nonumber \\
|\hat{\bm r} -\rangle &=& {1\over \sqrt{2}}
\left[
\begin{array}{r}
e^{-i\phi /2}\sin{\theta \over 2} \\
-e^{i\phi /2}\cos{\theta \over 2}
\end{array}
\right].
\label{r_pm}
\end{eqnarray}
Combining these two types of spinors, one can compose 
the spinorial part of $\psi$ that can be matched with the condition
(\ref{tau_z}), leading to
\begin{equation}
[m_\perp [\kappa] + \kappa A\alpha\beta]|\hat{\bm r} \alpha \rangle_\beta =0,
\label{radial_3}
\end{equation}
where
\begin{equation}
|\hat{\bm r} \alpha \rangle_\beta
=
|\hat{\bm r} \alpha \rangle
\otimes
|\tau_y \beta \rangle.
\end{equation}
Eq. (\ref{radial_3}) implies
\begin{equation}
\kappa
={-\alpha\beta A \pm \sqrt{A^2+4m_0 m_2}\over 2m_2}.
\label{k_pm}
\end{equation}
The radial part of the wave function $\rho (r)$
that is compatible with the boundary condition
(\ref{bc})
takes the form given in Eq. (\ref{psi_perp_A2}),
here, with $\kappa_\pm$ being the two solutions of Eq. (\ref{k_pm}).
Clearly, both $\kappa_+$ and $\kappa_-$ must be positive,
for the wave function $\psi$
to describe a surface state
localized in the vicinity of the spherical surface ($r \simeq R$).
Thus,
in order to cope with the boundary condition,
one must have both
$\alpha\beta <0$
and
$m_0 m_2 <0$
for the choice of model parameters such that
$A/m_2 > 0$.
Notice that the second condition, $m_0 m_2 <0$,
which has appeared here automatically from the boundary condition,
is a requirement for the system to be in the topologically 
non-trivial phase.

We have thus successfully found
the two basis eigenstates of $H_\perp$ for constructing
the effective surface Hamiltonian.
For simplicity of the notation we denote them as
$\psi = |\pm \rangle\rangle$,
where
\begin{eqnarray}
|+ \rangle\rangle &=& \rho (r) |\hat{\bm r} + \rangle_-,
\nonumber \\
|- \rangle\rangle &=& \rho (r) |\hat{\bm r} - \rangle_+.
\end{eqnarray}
To avoid misunderstanding of the notations
let us express explicitly
the four-component vectorial form of 
$|\hat{\bm r} \pm \rangle_\mp$
\begin{eqnarray}
|\hat{\bm r} + \rangle_- &=& {1\over 2}
\left[
\begin{array}{c}
\left[
\begin{array}{c}
1\\ -i
\end{array}
\right]
e^{-i\phi /2}\cos{\theta \over 2} \\
\left[
\begin{array}{c}
1\\ -i
\end{array}
\right]
e^{i\phi /2}\sin{\theta \over 2}
\end{array}
\right],
\nonumber \\
|\hat{\bm r} - \rangle_+ &=& {1\over 2}
\left[
\begin{array}{c}
\left[
\begin{array}{c}
1\\ i
\end{array}
\right]
e^{-i\phi /2}\sin{\theta \over 2} \\
-
\left[
\begin{array}{c}
1\\ i
\end{array}
\right]e^{i\phi /2}\cos{\theta \over 2}
\end{array}
\right].
\label{dv_basis}
\end{eqnarray}
Here, the arrangement of the basis is made in accordance with that
of Eq. (\ref{matH}).
Notice that the eigenvectors of Eq. (\ref{dv_basis}) are
{\it double-valued} with respect to the azimuthal angle $\phi$.
This does not happen for the polar angle $\theta$, since the domain of
definition for $\theta$ is restricted to a finite range, $\theta \in [0, \pi]$, 
and not periodic, in contrast to $\phi$.
The double-valuedness stems from our choice of the
(arbitrary) phase factor in front of Eq. (\ref{r_pm}).
This is, on the other hand, 
merely a choice, and one can equally formulate
the same problem consistently using a pair of {\it single-valued} eigenvectors.
We leave further arguments on this point to Sec. III and
here take these {\it double-valued} eigenvectors
as a basis for constructing the surface effective Hamiltonian.

The effective surface ''Hamiltonian'' ${\cal H}_{\rm dv}$ acts on a two-component spinor,
\begin{equation}
\bm\alpha=
\left[
\begin{array}{l}
\alpha_+\\
\alpha_-
\end{array}
\right].
\label{alpha1}
\end{equation}
Within the $k\cdot p$-approximation
any surface state $|\bm\alpha \rangle\rangle$ can be represented as a linear combination of
$|+ \rangle\rangle$ and $|- \rangle\rangle$
with the amplitude specified, respectively, by
$\alpha_+$ and $\alpha_-$, i.e.,
\begin{equation}
|\bm\alpha\rangle\rangle =\alpha_+ |+ \rangle\rangle + \alpha_- |- \rangle\rangle
\label{alpha2}
\end{equation}
and
\begin{equation}
\left[
\begin{array}{l}
\langle\langle +| H_\parallel |\bm\alpha\rangle\rangle \\
\langle\langle -| H_\parallel |\bm\alpha\rangle\rangle
\end{array}
\right]
\equiv  {\cal H}_{\rm dv} \bm\alpha.
\end{equation}
The explicit form of
${\cal H}_{\rm dv}$
can be determined by evaluating each of the matrix elements
$H_\parallel$ against the basis vectors $|\pm\rangle\rangle$,
i.e., 
$\langle\langle \pm| H_\parallel |\mp \rangle\rangle$.
The procedure we follow here is precisely in parallel with that of the standard
degenerate perturbation theory.
$H_0=H_\perp$ is an unperturbed Hamiltonian and $|\pm \rangle\rangle$
are its (two-fold) degenerate eigenstates.
To find the (degeneracy-lifted) spectrum of the perturbed Hamiltonian, 
$H_{\rm tot}=H_0 +H'$,
where $H' = H_\parallel$, $H_{\rm tot} = H_{\rm bulk}$
we first calculate the matrix elements: 
$\langle\langle \alpha| H' |\beta \rangle\rangle \equiv ({\cal H}_{\rm dv})_{\alpha\beta}$
($\alpha, \beta =\pm$),
then diagonalize this $2\times 2$ coefficient matrix.

The explicit matrix form of $H_\parallel$ is
\begin{equation}
H_\parallel =
\left[
\begin{array}{cccc}
m_\parallel & -i A D_z  & 0 & -i A D_- \\
-i A D_z  & - m_\parallel  & -i A D_- & 0 \\
0 & -i A D_+ & m_\parallel  & i A D_z \\ 
-i A D_+ & 0 & i A D_z & - m_\parallel   
\end{array}
\label{H_para}
\right],
\end{equation}
where
$D_\pm$ and $D_z$ are defined
in Eqs. (\ref{D_pm}) and (\ref{D_z}).
Performing the $r$-integral in
$\langle\langle \alpha| H_\parallel |\beta \rangle\rangle$
one can safely replace 
the $r$-dependence in these expressions
with the radius $R$ of the sphere,
assuming that 
the surface wave function is well localized in the vicinity of the surface.
Alternatively, one can equally regard
$D_\pm$ and $D_z$ as
\begin{equation}
D_\pm \simeq 
{e^{\pm i\phi}\over R}
 \left[ 
\cos \theta {\partial \over \partial \theta} 
\pm {i \over \sin \theta} {\partial \over \partial \phi} 
\right],\ 
D_z \simeq
- {\sin \theta \over R}{\partial \over \partial \theta}.
\end{equation}
At leading order in the expansion with respect to $1/R$
the diagonal terms of $H_\parallel$ can be neglected,
since 
\begin{equation}
m_\parallel =m_2{\bm L^2 \over r^2} \sim {1 \over R^2}
\end{equation}
[see also Eqs. (\ref{nabla}) and (\ref{m_perp})].
Within this accuracy
the coefficient matrix ${\cal H}_{\rm dv}$ is found,
after some algebra, to be
\begin{equation}
{\cal H}_{\rm dv}={A\over R}
\left[
\begin{array}{cc}
0  & -\partial_\theta + {i\partial_\phi\over \sin\theta} - {\cot\theta \over 2} \\
\partial_\theta + {i\partial_\phi\over \sin\theta} + {\cot\theta \over 2} & 0
\end{array}
\right].
\label{h2}
\end{equation}
Apart from an 
overall constant in front of the expression,
this can be identified as the ``Dirac operator'' for a free massless fermion
on the sphere.
\cite{Fukui, Eguchi, Kitsukawa,
Guinea_PRL, Guinea_NPB, Abrikosov_arxiv, Abrikosov, EPJB_fullerene}


The origin of the Berry phase term can be attributed to
the covariance of the derivatives
$\partial_\theta$ and 
$\partial_\phi$
on a curved spherical surface.
\cite{Guinea_PRL, Guinea_NPB, Abrikosov_arxiv, Abrikosov, EPJB_fullerene}
In this regard, the Berry phase term appears as a spin connection in ${\cal D}_\phi$ as
\begin{equation}
{\cal D}_\phi = \partial_\phi + {i \sigma_z\over 2} \cos \theta,
\end{equation}
replacing $\partial_\phi$ in Eq. (\ref{h2}) as
\begin{equation}
{\cal H}_{\rm dv} =
A \left[
\left( { -i \partial_\theta\over R} \right) \sigma_y 
- \left( { -i {\cal D}_\phi \over R \sin\theta} \right) \sigma_x 
\right].
\end{equation}
Alternatively,
the Berry phase term can be absorbed in $\partial_\theta$
by introducing
\begin{equation}
\tilde{\partial}_\theta = \partial_\theta + {1\over 2} \cot \theta.
\end{equation}
In terms of $\tilde{\partial}_\theta$,
the Dirac operator (\ref{h2}) 
can be also rewritten as ({\it cf.} Table I)
\begin{eqnarray}
{\cal H}_{\rm dv}&=&
{A \over R}
\left[
\begin{array}{cc}
0  & -\tilde{\partial}_\theta + {i\partial_\phi\over \sin\theta}\\
\tilde{\partial}_\theta + {i\partial_\phi\over \sin\theta}& 0
\end{array}
\right]
\nonumber \\
&=& A \left[
\left( { -i \tilde{\partial}_\theta\over R} \right) \sigma_y 
- \left( { -i \partial_\phi \over R \sin\theta} \right) \sigma_x 
\right].
\label{h2_v2}
\end{eqnarray}

\section{Nature of the two types of Berry phase}

The advantage of considering the spherical geometry is that
the existence of two 
different types of Berry phase becomes apparent;
each associated, respectively, with
an electonic motion in the polar ($\theta$-) [type (A)] 
and azimuthal ($\phi$-) [type (B)]
directions (see Table \ref{table_Berry}).
The type (A) Berry phase is intrinsic to the curvature of the spherical surface,
while the type (B) is 
associated with the so-called spin-to-surface locking .
On a cylindrical surface, on contrary,
only the latter [type (B)] manifests,
since the cylindrical surface has a vanishing Gaussian curvature.
The contrasting behaviors of the two types Berry phase are
summarized in Table \ref{table_Berry}.

To highlight the distinct behaviors of the two types of Berry phase,
let us reconsider
the Dirac operator (\ref{h2})
expressed against the double-valued basis vectors
(\ref{dv_basis}).
As mentioned earlier,
this was not a unique choice of the basis.
One can equally choose
them
to be {\it single-valued} as
\begin{eqnarray}
|+ \rangle\rangle &=& {1\over 2}
\left[
\begin{array}{l}
\left[
\begin{array}{c}
1\\ -i
\end{array}
\right]
\cos{\theta \over 2} \\
\left[
\begin{array}{c}
1\\ -i
\end{array}
\right]
e^{i\phi}\sin{\theta \over 2}
\end{array}
\right],
\nonumber \\
|- \rangle\rangle &=& {1\over 2}
\left[
\begin{array}{l}
\left[
\begin{array}{c}
1\\ i
\end{array}
\right]
\sin{\theta \over 2} \\
-
\left[
\begin{array}{c}
1\\ i
\end{array}
\right]
e^{i\phi}\cos{\theta \over 2}
\end{array}
\right].
\label{sv_basis}
\end{eqnarray}
This type of a single-valued choice of the basis is often employed
in the $\bm k\cdot\bm p$-description of the electronic states in graphene.
\cite{TAndo}
Once this choice of basis vectors is adopted,
one can repeat the same procedure as we described in the last section,
to find the surface effective Hamiltonian ${\cal H}_{\rm sv}$, 
or the Dirac operator in this basis, as
\begin{equation}
{\cal H}_{\rm sv}={A\over R}
\left[
\begin{array}{cc}
0  & -\partial_\theta + {i\partial_\phi\over \sin\theta} - {1\over 2} \cot {\theta \over 2} \\
\partial_\theta + {i\partial_\phi\over \sin\theta} - {1\over 2} \tan {\theta \over 2} & 0
\end{array}
\right].
\label{h1}
\end{equation}
In passing from Eq. (\ref{h2}) to (\ref{h1}),
the matrix elements are replaced as
\begin{equation}
-i \partial_\phi \rightarrow -i \partial_\phi +{1\over 2}.
\label{replace}
\end{equation}
Here, the additive factor $1/2$
is nothing but the Berry phase $\pi$ associated with the spin-to-surface locking
in cylindrical surfaces that has been extensively discussed in the literature.
\cite{Ashvin_PRL, Mirlin_PRL, JMoore_PRL, k1, k2}

The appearance of the type (A) Berry phase (see Table \ref{table_Berry})
is not restricted to the topological insulator 
surface state.
It has already appeared in the study of the electronic spectrum of fullerene,
typically the one called, ``buckyball'' (or Buckminsterfullerene).
\cite{Guinea_PRL, Guinea_NPB, Abrikosov_arxiv, Abrikosov, EPJB_fullerene}
The type (B) Berry phase, on contrary, is specific to the topological insulator 
surface state.
By its nature whether this type of Berry phase appears explicitly in the effective
Dirac operator depends on the choice of the basis.
In ${\cal H}_{\rm dv}$ [Eq. (\ref{h2})] the Berry phase is superficially
hidden in the anti-periodicity of the basis spinor (\ref{dv_basis})
with respect to the azimuthal angle $\phi$.
When one considers the orbital part of the wave function or spinor (\ref{alpha1}),
this point must be carefully taken into account in its periodicity
with respect to $\phi$.
This point will be clarified
in the next section.

The Berry phase term, or  more precisely the spin connection
of the form of $\pm \cot \theta /2$ in Eq. (\ref{h2}),
or equivalently,
either $\cot (\theta /2)/2$ or $\tan (\theta /2)/2$ 
in the two off-diagonals of Eq. (\ref{h1}),
can be interpreted as
a vector potential generated by an effective {\it magnetic monopole}.
Indeed,
the magnetic field associated with a magnetic monopole of
strength $g$ can be successfully encoded in a vector potential,
e.g.,
\begin{eqnarray}
\bm A_{I} = {g \over 4\pi r} \tan {\theta \over 2}\ \hat{\bm \phi},
\\
\bm A_{II} =  {-g \over 4\pi r} \cot {\theta \over 2}\ \hat{\bm \phi},
\end{eqnarray}
by introducing the concept of Dirac's string.
Here, $\bm A_{I}$ and $\bm A_{II}$ correspond to a choice of the gauge in which
the Dirac's string runs, respectively, on the $-z$- ($+z$-) axis 
[direction of the south ($\theta =\pi$) vs. north ($\theta =0$) poles].
Eqs. (\ref{h2}), (\ref{h1}) imply
that the strength of the ``induced'' monopole is, respectively,
$g = -2\pi$ for $\alpha_+$, and 
$g = 2\pi$ for $\alpha_-$.
A fictitious magnetic monopole of an {\it opposite} charge is
effectively induced at the center of the sphere
for the two spin components of the surface spinor wave function (\ref{alpha1})
[see also Eq. (\ref{q})].

\section{Surface eigenstates on the sphere}

To find the electronic spectrum of the spherical topological insulator surface state,
one needs to solve the surface Dirac equation,
\begin{equation}
{\cal H}_{\rm dv} \bm \alpha = E_\parallel \bm \alpha
\label{dv_eigen}
\end{equation}
explicitly.
In accordance with the decomposition of the Hamiltonian:
$H_{\rm bulk} = H_\perp + H_\parallel$, we have also separated 
the energy eigenvalue $E$ of the original 
eigenvalue equation,
\begin{equation}
H_{\rm bulk} 
|\bm \alpha \rangle\rangle
= E |\bm \alpha \rangle\rangle
\label{alpha3}
\end{equation}
into two parts: $E = E_\perp + E_\parallel$.
But, of course, since $E_\perp =0$ [Eq. (\ref{zero})],
$E=E_\parallel$.
As we have already emphasized, there exists some freedom
in the choice of basis vectors, but here 
mainly to ease comparison with the literature
we take them double-valued
as in Eq. (\ref{dv_basis}).
To solve the eigenvalue equation (\ref{dv_eigen}) explicitly
we first separate the variables as
$\bm \alpha (\theta, \phi) = e^{ im\phi} \bm \alpha_m (\theta)$, 
or in terms of the components,
\begin{equation}
\left[
\begin{array}{l}
\alpha_+ (\theta, \phi) \\
\alpha_- (\theta, \phi)
\end{array}
\right]
= e^{ im\phi}
\left[
\begin{array}{l}
\alpha_{m+} (\theta) \\
\alpha_{m-} (\theta)
\end{array}
\right].
\end{equation}
Clearly, the quantum number $m$ signifies physically 
the $z$-component of the orbital angular momentum.
Here, 
the important point is that
this $m$ takes {\it half-odd integral} values,
\begin{equation}
m=\pm {1 \over 2}, \pm {3 \over 2}, \pm {5 \over 2}, \cdots
\label{m_half}
\end{equation}
in contrast to the case of fullerene.
\cite{Guinea_PRL, Guinea_NPB, EPJB_fullerene}
Any solution of the original Scr\"odinger equation $H_{\rm bulk} |\psi \rangle = E |\psi \rangle$,
or its surface solution $|\bm \alpha \rangle\rangle$ [see Eq. (\ref{alpha2})]
obeys to a periodic boundary condition with respect to $\phi$.
To be explicit, any bulk (or surface) solution
$|\psi \rangle =|\psi (r, \theta, \phi) \rangle$
satisfies $| \psi (r, \theta, \phi +2\pi) \rangle = |\psi (r, \theta, \phi) \rangle$,
and in the same sense,
\begin{equation}
|\bm \alpha (\theta, \phi +2\pi) \rangle\rangle = |\bm \alpha (\theta, \phi) \rangle\rangle.
\label{pbc}
\end{equation}
On the other hand,
the basis spinor (\ref{dv_basis}) spanning the space of the surface solutions 
$|\bm \alpha (\theta, \phi) \rangle\rangle$
[Eq. (\ref{alpha2})]
is anti-periodic with respect to $2\pi$-rotation of the azimuthal angle,
i.e.,
$|\pm \rangle\rangle \rightarrow - |\pm \rangle\rangle$
under $\phi \rightarrow \phi + 2\pi$.
In order to insure the periodicity of the total wave function (\ref{pbc})
this minus sign must be compensated, or absorbed in the prefactor
$\alpha_\pm = \alpha_\pm (\theta, \phi)$ of Eq. (\ref{alpha2}),
i.e., $\alpha_\pm (\theta, \phi + 2\pi) = - \alpha_\pm (\theta, \phi)$

In terms of $\alpha_{m+} (\theta)$ and $\alpha_{m-} (\theta)$,
the eigenvalue equation (\ref{dv_eigen}) becomes a couple of
first-order linear differential equation,
\begin{eqnarray}
-{A \over R}
\left[
{d\over d\theta} + {m \over \sin\theta} + {\cot\theta \over 2}
\right] \alpha_{m-} (\theta)
&=& E_\parallel \alpha_{m+} (\theta),
\nonumber \\
{A \over R}
\left[
{d\over d\theta} - {m \over \sin\theta} + {\cot\theta \over 2}
\right] \alpha_{m+} (\theta)
&=& E_\parallel \alpha_{m-} (\theta).
\label{couple}
\end{eqnarray}
These two equations combine to give
\begin{eqnarray}
\Big[
{1\over \sin\theta} {d\over d\theta} \sin\theta {d\over d\theta} 
&-& {1\over \sin^2 \theta} \left( m-{\sigma\over 2}\cos\theta \right)^2
\nonumber \\
&+& \lambda^2 - {1 \over 2}
\Big] \alpha_{m\sigma} (\theta) = 0,
\label{Abr_19}
\end{eqnarray}
in which two important parameters $\sigma$ and $\lambda$
have been introduced.
$\sigma =\pm$ specifies the spin index in the subscript of $\alpha_{m\sigma}$,
whereas $\lambda$ parametrizes the energy as
\begin{equation}
E_\parallel = {A \over R}\lambda.
\label{E_para}
\end{equation}

Let us remark here that
Eq. (\ref{Abr_19}) 
is formally equivalent to a differential equation
defining the so-called monopole harmonics, \cite{WuYang}
the latter describing (the angular part of)
the electronic motion
in the presence of a magnetic monopole.
The role of the non-trivial Berry curvature
imposed by the spherical geometry
can be thus interpreted
as an effective {\it magnetic monopole} induced at the center of the sphere.
An electron in the surface state of a spherical topological insulator
''sees'' such an effective magnetic monopole and behaves accordingly.
In the notation of Ref. \cite{WuYang},
the quantum number $q$ specifies physically 
the strength $g$ of the effective 
magnetic monopole placed at the origin as
$g= 4\pi q$.
The Dirac's quantization condition restricts
the ''allowed'' value of this quantum number $q$
to be half integral:
$q=0, \pm1/2, \pm 1, \cdots$.
Here, in Eq. (\ref{Abr_19}), $q$ is identified as
\begin{equation}
q= - {\sigma\over 2} = \pm {1\over 2}.
\label{q}
\end{equation}
The eigenfunction $\alpha_{m\sigma} (\theta)$ of Eq. (\ref{Abr_19}) 
is indeed related to the monopole harmonics specified by this value of $q$.
Thus at the center of a spherical topological insulator
a magnetic monopole of strength $\pm 2\pi$, 
the smallest finite value compatible with the Dirac's quantization condition
is effectively induced.
By its nature the induced monopole is automatically regularized by a Dirac's string.

Introducing a new independent variable
$\zeta=\cos \theta$,
one can rewrite Eq. (\ref{Abr_19}) as
\cite{Abrikosov_arxiv}
\begin{equation}
\left[
{d \over d\zeta} (1-\zeta^2) {d \over d\zeta}
- {m^2-\sigma m \zeta + 1/4 \over 1-\zeta^2}
+ \lambda^2 - {1 \over 4}
\right] \alpha_{m\sigma} = 0.
\label{Abr_20}
\end{equation}
This can be further rewritten in the form of Jacobi-type
differential equation.
As shown below,
its normalizable solutions are known to be
expressed in terms of Jacobi Polynomials $P_n^{\mu\nu} [\zeta]$
(see Appendix C for our conventions).
Changing the dependent variables as
\begin{equation}
\alpha_{m\sigma} [\zeta] =
(1 - \zeta)^{{1\over2} |m - {\sigma\over2}|} (1 + \zeta)^{{1\over2} |m + {\sigma\over2}|} \beta_{m\sigma} [\zeta],
\label{beta1}
\end{equation}
and using the fact that $m$ is half-integral [Eq. (\ref{m_half})],
one can verify,
\begin{eqnarray}
&&\Big[
(1-\zeta^2) {d^2 \over d\zeta^2}
+ \left\{ \sigma{m\over |m|} - (2|m|+2) \zeta \right\} {d\over d\zeta}
\nonumber \\
&& - |m| (|m| +1) + \lambda^2 - {1 \over 4}
\Big] 
\beta_{m\sigma} [\zeta] = 0
\label{Abrikosov_22}
\end{eqnarray}
Comparing this with the standard form of the Jacobi's differential equation
({\it c.f.} Eq. (\ref{hyper})),
\begin{eqnarray}
&&\Big[
(1-\zeta^2) {d^2 \over d\zeta^2}
+ \left\{ \nu-\mu - (\mu +\nu +2) \zeta \right\} {d\over d\zeta}
\nonumber \\
&&+ n(n +\mu + \nu +1)
\Big] P_n^{\mu\nu} [\zeta]=0,
\label{Jac_D}
\end{eqnarray}
one can identify the parameters as
\begin{eqnarray}
\mu &=& |m| - {\sigma \over 2}{m\over |m|} = \left| m - {\sigma \over 2} \right|,
\nonumber \\
\nu &=& |m| + {\sigma \over 2}{m\over |m|} = \left| m + {\sigma \over 2} \right|,
\label{munu}
\end{eqnarray}
and
\begin{eqnarray}
\lambda^2 &=& n(n+\mu +\nu +1)+  |m| (|m| +1) + {1 \over 4}
\nonumber \\
&=& \left( n +|m| + {1\over 2} \right)^2,
\label{lambda}
\end{eqnarray}
where the normalizability of the wave function requires $n$ to be non-negative integers,
\begin{equation}
n = 0, 1, 2, 3, \cdots.
\label{n_int}
\end{equation}
In Eqs. (\ref{munu}) $\sigma =\pm$ referes to the subscript of $\beta_{m\sigma} [\zeta]$.
The second equality of Eqs. (\ref{munu}) holds since
$m$ is half-integral.
Thereby,
apart from a normalization constant $c_{nm\sigma}$, 
which will be determined later,
the surface wave function $\beta_{m\sigma} [\zeta]$ is expressed
in terms of the $n$-th order Jacobi polynomial as
\begin{equation}
\beta_{m\sigma} [\zeta] = \beta_{nm\sigma} [\zeta] 
= c_{nm\sigma} P_n^{|m - {\sigma \over 2}|,|m + {\sigma \over 2}|} [\zeta].
\label{beta2}
\end{equation}
Clearly, $n$ is common to $\beta_{m+} [\zeta]$ and $\beta_{m-} [\zeta]$
once $\lambda$ is chosen to be an appropriate quantized value imposed by
Eq. (\ref{lambda}). 
Substituting Eq. (\ref{lambda}) into (\ref{E_para}), one finds the surface energy spectrum,
\begin{equation}
E=E_\parallel = \pm {A \over R} \left( n +|m| + {1\over 2} \right)
\equiv E_\lambda,
\label{spectrum}
\end{equation}
where $m$ and $n$ take, respectively,
half-integral and non-negative integral values
[see Eqs. (\ref{m_half}) and (\ref{n_int})].
The energy spectrum (\ref{spectrum}) has a couple of specific features:
\begin{enumerate}
\item
The discrete energy levels $E_\lambda$ are placed with an equal distance,
$A/R$, symmetrically on the positive and negative side of $E$,
excluding the zero energy, $E=0$.
\item
The degeneracy $g_\lambda$ of each energy level
increases linearly with $|E|$,
taking every (positive) even numbers,
$g_\lambda = 2, 4, 6, \cdots$,
when one starts counting it
at the first positive and negative energy level with
$\lambda =\pm 1$.
\end{enumerate}

The discrete energy levels $E_\lambda$
can be also casted from the view of finite-size energy gap.
On the surface of an infinitely long cylinder
the surface electronic spectrum shows an energy gap
as a consequence of the spin-to-surface locking and
the resulting half-integer quantization
analogous to Eq. (\ref{m_half}).
The size of the obtained energy gap
on a cylindrical surface is inversely proportional 
to the radius of the cylinder.
Here, in Eq. (\ref{spectrum})
the zero energy $E=0$ is indeed excluded,
which can be regarded as a remnant of the energy gapped 
in the continuos spectrum.
Note that this has nothing to do with the discreteness of
the spectrum due to finite-size ``quantization''.
What counts here is the absolute position of the entire (discrete) spectrum.
The size of the ``energy gap'',
\begin{equation}
\Delta (R) = E_1 - E_{-1} = {2A \over R},
\end{equation}
is again inversely proportional $R$, the radius of the sphere.
\cite{Guinea_2011}


In order
to determine the relative magnitude and phase of
$c_{nm+}$ and $c_{nm-}$,
one needs to go back to Eqs. (\ref{couple})
and substitute $\beta_{m\sigma} [\zeta]$
obtained as in Eqs. (\ref{beta2})
into this couple of equations.
This is straight-forward,
but turns out to be rather a tedious work.
Leaving explicit demonstration of this to Appendix D,
here we refer only to its result:
\begin{equation}
c_{nm-}= - {\rm sign} [\lambda m] c_{nm+}.
\label{cnm1}
\end{equation}
Choosing $c_{nm+}$ to be real, 
and taking the normalization,
\begin{equation}
2\pi \int d(\cos \theta) |\bm \alpha_{nm} (\theta)|^2 =1,
\label{norm}
\end{equation}
also into account ({\it c.f.} Eq. (\ref{Jac_norm}) for the normalization of $P_n^{\mu\nu} [\zeta]$),
one can give an explicit list of the coefficients $c_{nm\sigma}$,
in which
\begin{equation}
c_{nm+} = {1\over \sqrt{2\pi}} {\sqrt{n!(n+2|m|)!} \over 2^{|m|+1/2} (n+|m|-1/2)!}.
\label{cnm2}
\end{equation}
In Eq. (\ref{norm}) we introduced the notation $\bm \alpha_{nm} (\theta)$, i.e.,
$\bm \alpha_m (\theta) = \bm \alpha_{nm} (\theta)$
in accordance with Eq. (\ref{beta2}).


As a solution of the effective surface Dirac equation (\ref{dv_eigen}),
Eqs. (\ref{cnm2}), (\ref{cnm1}), (\ref{beta2}), (\ref{beta1})
specify an explicit form of the surface eigenspinor,
$\bm \alpha (\theta, \phi) = e^{i m\phi} \bm \alpha_{nm} (\theta)$,
with a definite {\it half-integral} angular momentum $m$.
It may be suggestive to give a few concrete examples of
$\bm \alpha_{nm} (\theta)$ for small values of $n$ and $|m|$.
For simplicity
let us restrict ourselves to the case of $\lambda >0$ (positive energy).
For $n=0$ with an arbitrary half-integral $m$,
$\bm \alpha_{nm} (\theta)$ is found to be
\begin{eqnarray}
&&\bm \alpha_{0m} (\theta) =
\\
&&{1\over \sqrt{4\pi}}
{\sqrt{(2|m|)!} \over (|m|-{1\over2})!}
\left[
\begin{array}{r}
\left( \sin {\theta\over 2} \right)^{|m-{1\over 2}|}
\left( \cos {\theta\over 2} \right)^{|m+{1\over 2}|}
\\
- s_m
\left( \sin {\theta\over 2} \right)^{|m+{1\over 2}|}
\left( \cos {\theta\over 2} \right)^{|m-{1\over 2}|}
\end{array}
\right],
\nonumber
\end{eqnarray}
where
$s_m$ is an abbreviated notation for
\begin{equation}
s_m = {\rm sign} [m] = {m \over |m|}.
\end{equation}
More specifically,
\begin{eqnarray}
\bm \alpha_{0{1\over 2}} &=&
\left[
\begin{array}{l}
\alpha_{0{1\over 2}+} \\
\alpha_{0{1\over 2}-}
\end{array}
\right]
={1\over\sqrt{4\pi}}
\left[
\begin{array}{r}
\cos {\theta\over 2} \\
-\sin {\theta\over 2}
\end{array}
\right],
\label{a01}
\\
\bm \alpha_{0, -{1\over 2}} &=&
{1\over\sqrt{4\pi}}
\left[
\begin{array}{r}
\sin {\theta\over 2} \\
\cos {\theta\over 2}
\end{array}
\right],
\label{a0m1}
\\
\bm \alpha_{0{3\over 2}} &=&
\sqrt{3\over 2\pi}
\left[
\begin{array}{r}
\sin {\theta\over 2} \cos^2 {\theta\over 2} \\
- \sin^2 {\theta\over 2} \cos {\theta\over 2}
\end{array}
\right],
\label{a03}
\\
\bm \alpha_{0, -{3\over 2}} &=&
\sqrt{3\over 2\pi}
\left[
\begin{array}{r}
\sin^2 {\theta\over 2} \cos {\theta\over 2} \\
\sin {\theta\over 2} \cos^2 {\theta\over 2}
\end{array}
\right].
\label{a0m3}
\end{eqnarray}
For $n=1$ and $m=\pm 1/2$,
\begin{eqnarray}
\bm \alpha_{1{1\over 2}} &=&
- {1\over 2\sqrt{2\pi}}
\left[
\begin{array}{r}
\cos {\theta\over 2} (1-3\cos\theta)\\
\sin {\theta\over 2} (1+3\cos\theta)
\end{array}
\right],
\label{a11}
\\
\bm \alpha_{1, -{1\over 2}} &=&
{1\over 2\sqrt{2\pi}}
\left[
\begin{array}{r}
\sin {\theta\over 2} (1+3\cos\theta)\\
- \cos {\theta\over 2} (1-3\cos\theta)
\end{array}
\right].
\label{a1m1}
\end{eqnarray}
The two eigenstates (\ref{a01}) and (\ref{a0m1})
constitute the two lowest energy (degenerate) eigenstates on the $E>0$ side,
whereas Eqs. (\ref{a03}-\ref{a1m1}) correspond to the four
second lowest (first excited) states.

\begin{figure}
\begin{center}
\includegraphics[width=8cm]{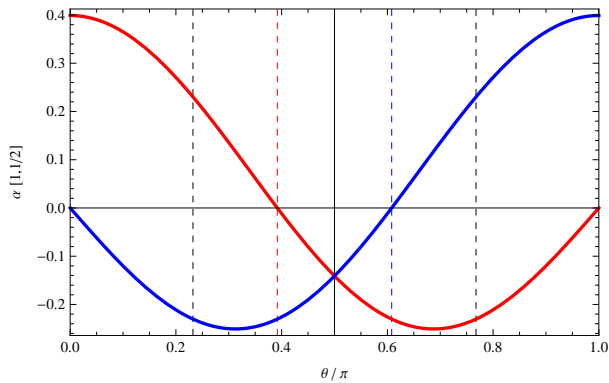}
\end{center}
\caption{$\theta$-dependence of the two spin components of
$\bm \alpha_{1{1\over 2}} (\theta)$.
$\alpha_{1{1\over 2}+} (\theta)$
and
$\alpha_{1{1\over 2}-} (\theta)$, shown, respectively, in red and in blue.
The two spin components acquire an equal weight at $\theta=\theta_1 \simeq 0.232 \pi$,
$\theta=\pi /2$
and at $\theta = \pi - \theta_1$.
$\alpha_{1{1\over 2}+} (\theta)$ vanishes at
$\theta = \theta_2 \simeq 0.392 \pi$ and at $\theta = \pi$, whereas
$\alpha_{1{1\over 2}-} (\theta)$ vanishes at $\theta = 0$
and at $\theta = \pi - \theta_2$.}
\label{plot_wf}
\end{figure}


Let us point out that the surface spinor wave functions thus constructed
imply a rich {\it spin texture} that the surface eigenstates manifest.
This can be already seen in the simplest example,
the case of the eigenstate $\bm \alpha_{0{1\over 2}}$ given as in Eq. (\ref{a01}).
At the north pole ($\theta =0$)
the spin in this eigenstate is
pointed in the $+\bm z$-direction,
perpendicular to the tangential plane of the sphere at this point.
As $\theta$ increases, the spin tends to lie closer to the tangential plane.
At $\theta =\pi /2$, i.e., on the equator the spin is completely in-plane to the spherical surface.
Now, staying at $\theta =\pi /2$, 
if one let $\phi$ vary from $0$ to $2\pi$, i.e.,
as the electron hypothetically travels around the equator,
the spin also completes a $2\pi$ rotation in the $(x,y)$-plane,
manifesting the feature of ``spin-to-surface locking''.
As one further increases $\theta$, the spin down component starts to dominate,
before it dominates completely at the south pole.


Another example showing a more complicated spin texture
is the case of $\bm \alpha_{1{1\over 2}} (\theta)$
given as in Eq. (\ref{a11})
[the two spin components of $\bm \alpha_{1{1\over 2}} (\theta)$ are
depicted in FIG. \ref{plot_wf}].
At the north pole the spin in this eigenstate is again
pointed in the $+\bm z$-direction.
As $\theta$ increases, it starts to lie, but tends to lie
more strongly than the case of $\bm \alpha_{0{1\over 2}}$.
At 
\begin{equation}
\theta = 2\arccos {3+\sqrt{5}\over 6} \simeq 0.232\pi
\equiv \theta_1,
\end{equation}
the two spin components acquire an equal weight.
At this angle $\theta=\theta_1$,
if one let $\phi$ vary from $0$ to $2\pi$,
complete spin-to-surface locking occurs.
As $\theta$ exceeding $\theta_1$,
the $|\hat{\bm r}-\rangle$-component starts dominate.
At 
\begin{equation}
\theta=\arccos [1/3] \simeq 0.392 \pi
\equiv \theta_2,
\end{equation}
the centrifugal spin component
$\alpha_{1{1\over 2}+} (\theta)$ vanishes.
Therefore,
at this point
the $|\hat{\bm r}-\rangle$-component dominates completely,
and the spin is pointed to the center of the sphere.
As $\theta$ further increases,
the spin gradually tilts back to the tangential plane
(at $\theta=\pi/2$, spin-to-surface locking is recovered),
then it starts to be further tilted toward the outside of the sphere.
At $\theta=\arccos [-1/3] =\pi -\theta_2$,
the centripetal spin component 
$\alpha_{1{1\over 2}-} (\theta)$ vanishes,
and the spin finds itself purely in the $|\hat{\bm r}+\rangle$ state.
At $\theta > \pi - \theta_2$,
$|\hat{\bm r}-\rangle$ starts to dominate again, and
the spin is finally pointed in 
$+\bm z$-direction at the south pole.
The behavior of the spin as varying $\phi$ from $0$ to $2\pi$ is 
the same as the case of $\bm \alpha_{0{1\over 2}}$
since the two states share the same quantum number $m=1/2$.

The spin rotates more drastically 
also in the $\phi$-direction for the eigenstates 
with $|m| \ge 3/2$.
Clearly,
the surface spinor wave functions with quantum numbers $n$ and $|m|$ higher than
the examples given in Eqs. (\ref{a01}-\ref{a1m1})
show
a richer spin structure on the spherical surface
as a consequence of the interplay between the two types of Berry phase;
{\it c.f.} the case of cylindrical geometry in which only a single type
Berry phase manifests.

\begin{figure}
\begin{center}
\includegraphics[width=8cm]{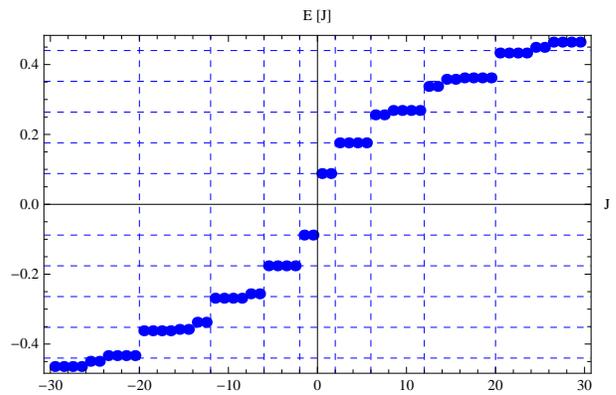}
\end{center}
\caption{Pseudo-degenerate spectrum $E [J]$ of the
surface eigenstates,
obtained by diagonalization of
the tight-binding Hamiltonian on a cubic lattice.
The system has a cubic shape of linear dimension $L =16$. $m_0/m_2= -1$, $A=1$.}
\label{spec_cube}
\end{figure}

\section{Comparison with the tight-binding calculation}

We have so far investigated specific features of the topological insulator surface state
occupying a finite volume,
taking as an example the case of spherical geometry.
Starting with the gapped bulk effective Hamiltonian,
we have derived and solved the surface Dirac equation,
from which we have deduced the surface electronic spectrum (\ref{spectrum})
and the explicit form of the spinor wave functions
[as given through Eqs. (\ref{beta1}, \ref{beta2}, \ref{cnm1}, \ref{cnm2})]
on a perfect spherical surface.
The role of two distinct types of Berry phase has been revealed.
Here, we take another viewpoint;
namely,
we go back to the bulk effective Hamiltonian [Eq. (\ref{H_bulk})],
and implemented it as a nearest-neighbor
tight-binding model,
which allows for obtaining the spectrum of
the surface solutions
by exact diagonalization.
We show that
the basic features on the surface energy spectrum
we have found so far
in the idealized spherical geometry with exact rotational symmetry
is {\it still valid}
when that symmetry is {\it weakly broken}.

Let us employ the following
lattice implementation of $H_{\rm bulk}$, i.e., Eq. (\ref{H_bulk})
on a cubic lattice:
\begin{equation}
H_{\rm lattice} = \tau_z m(\bm k) + A \tau_x \sum_{j=x,y,z} \sigma_j \sin k_j,
\label{H_lattice}
\end{equation}
where
\begin{equation}
m(\bm k) = m_0 + 2\sum_{j=x,y,z} m_2 (1-\cos k_j),
\label{mass_lattice}
\end{equation}
is a lattice version of Eq. (\ref{mass_conti}).
The model specified by Eqs. (\ref{H_lattice}) and (\ref{mass_lattice}) can be
regarded as a tight-binding model with only the nearest neighbor hopping.
This couple of equations determine the structure of 
energy bands
over the entire Brillouin zone, 
which also reproduces,
in the vicinity of the $\Gamma$ point,
the bulk effective Hamiltonian, $H_{\rm bulk}$ [Eq. (\ref{H_bulk})]
in the continuum limit.
The system we consider here
has a cubic shape of linear dimension $L$,
in which
the lattice points
placed with a unit lattice spacing are restricted to
\begin{equation}
1\le x, y, z \le L. 
\end{equation}

The obtained surface energy spectrum $E=E[J]$ is shown in FIG. \ref{spec_cube}.
Here, $J$ is an index for numbering each surface eigenstate with increasing
order of $E$.
For an aesthetic reason, and also to ease counting of the degree of (pseudo-) degeneracy
we have chosen this index $J$ to be half integral,
$J=\pm 1/2, \pm 3/2, \cdots$.
Numbering of the eigenstate is done in such a way that
starting with $E=0$
$J=1/2, 3/2, 5/2, \cdots$ is attributed to each eigenstate 
with increasing (decreasing) order of $E$ on the positive- (negative-) $E$ side
[i.e., a positive (negative) $J$ corresponds to a
positive- (negative-) energy level].
Depending on the number of lattice sites contained in the system
($=L^3$) and on the value of model parameters,
certain numbers of states appear in the bulk energy gap.
Those surface eigenstates of relatively small $|E|<m_0$
are also characterized by the
spatial profile of the corresponding wave function;
their wave function is localized in the vicinity of the cubic surface.
In this demonstration,
the system's size is $L=16$, 
and the model parameters are chosen as
$m_0/m_2= -1$, $A=1$.

$E=E[J]$ plotted in FIG. \ref{spec_cube} shows
a ``pseudo-degenerate'' feature
much reminiscent of the quantization characteristic on the spherical surface;
the spectrum (\ref{spectrum}),
and the degeneracy rule [Eqs. (\ref{n_int}) and  (\ref{m_half})].
Notice that the zero energy state $E=0$
is clearly excluded.
Horizontal gridlines are located at the positions of $E$ that are 
an integer multiple of $E [1]$, the first (positive) energy level.
These gridlines are shown for verifying 
that the energy levels are equally placed,
one of the characteristic features of the spectrum on the spherical surface.
Small deviation from the ``expected'' spectrum (\ref{spectrum})
can be seen for $|\lambda|>3$,
which can be interpreted as a consequence 
of the breaking of spherical symmetry.
Vertical gridlines are drawn for highlighting the degree of
the pseudo-degeneracy of each level.
If one recalls that a magnetic monopole is effectively induced
at the center of the sphere,
the (pseudo-) equally spaced spectrum 
illustrated in
FIG. \ref{spec_cube}
can be interpreted as Landau levels.
For $L=16$ the value of $E [1]$ is found to be $E [1] \simeq 0.0880$
in units of $\hbar A/a$, where
$A$ is the group velocity of the surface state,
and $a$ is the lattice constant.
Taking experimentally realistic values for $A$ and $a$,\cite{Liu_PRB}
the characteristic energy scale
$\hbar A/a$ is estimated to be on the order of 0.1-1 eV.

\section{Concluding remarks}

The protected surface state of topological insulator has an ''active'' property
that it reveals
only when it is embedded onto a curved surface.
On a cylindrical surface, it induces an effective magnetic solenoid of total flux $\pm \pi$
In the same sense, 
a magnetic monopole of strength $\pm 2\pi$ 
is induced
when it is embedded onto a sphere.
In this paper
we have explicitly examined this active property of the topological insulator surface state,
focusing on a most suggestive case of the spherical surface.
As a result, the following unique profile of such surface states has been found.
The two important features are
\begin{enumerate}
\item
a unique quantization rule; equally spaced spectrum
with the exception of the ``evaporated'' zero energy state, and the simple
degeneracy rule,
\item
a rich spin texture resulting from the nature of complicated spinor wave function.
\end{enumerate} 
These characteristic features derived analytically using an idealized spherical geometry
is then contrasted with a tight-binding calculation on a cubic lattice of cubic shape.
Unexpectedly profound agreement of the two results
suggests that
those features which we have demonstrated on a perfect sphere
capture the essential characteristic of the surface states occupying a finite volume
of more generic shape, inevitably involving a curved surface.
In this sense it is natural to expect that
the obtained spectrum be applicable to the spectroscopy
of topological insulator {\it nano-particles}.
In particular, we predict a unique photo absorption/emission spectrum
resulting from the equally spaced
energy levels of the
low-energy surface eigenstates.

Most of the existing works characterizing the topological insulator surface states
are based on Bloch states.
Topological properties, however, do not depend on the translational symmetry.
Here, we have demonstrated this, focusing on the angular momentum
\cite{Congjun1, Congjun2, Congjun3}
(instead of the linear momentum)
as the good quantum number.

\acknowledgments
KI acknowledges Akihiro Tanaka for valuable comments on the manuscript.
The authors are supported by KAKENHI; KI and TF by the ``Topological Quantum Phenomena'' [Nos. 23103511 and 23103502], 
and YT by a Grant-in-Aid for Scientific Research (C) [No. 24540375].

\appendix

\section{Symmetry class and topological invariants}

In the classification of
the Dirac Bogoliubov-de Gennes type Hamiltonian
in terms of the
time-reversal ($\Theta$), particle-hole ($\cal C$) and chiral ($\Gamma_5$) symmetries,
\cite{Zirnbauer, Schnyder_PRB, Kitaev_AIP, Schnyder_AIP, Ryu_NJP, TeoKane}
our starting bulk effective Hamiltonian (\ref{H_bulk})
falls on the class of AII,
to which $\mathbb{Z}_2$
topological insulators in two (2D) and three spatial (3D) dimensions are
classified.
This class of models has the symmetry,
$\Theta^2=-1$, ${\cal C}^2=0$ and $\Gamma_5 =0$
(here, ``$0$'' indicates that
the system does not possess that type of symmetry),
and are characterized by
$\mathbb{Z}_2$-type
bulk toploogical invariants
both in 2D \cite{KaneMele_Z2}
and 3D.
\cite{FuKaneMele, MooreBalents, Roy}
For the specific choice, $\epsilon (\bm p) =0$,
this symmetry is upgraded to the class DIII,
yielding
$\Theta^2=-1$, ${\cal C}^2=1$ and $\Gamma_5 =1$,
where for the specific Hamiltonian (\ref{H_bulk})
${\cal C}$ and $\Gamma_5$ are given by
${\cal C}=\sigma_y \tau_y K$ and
$\Gamma_5 =\tau_y$.
This symmetry class obeys to a $\mathbb{Z}$-type
bulk topological classification,
characterized by a winding number $\cal N$ for the Wilson-Dirac-like operator
in three dimensions.

In the following, 
we explicitly construct and evaluate this $\mathbb{Z}$-type
winding number $\cal N$.
Here, to ease the notation we rewrite Eq. (\ref{H_bulk})
in the specific case of $\epsilon (\bm p) =0$ as 
\begin{alignat}1
H = A \tau_1\sigma_\mu p_\mu + \tau_3 m (\bm p),
\end{alignat}
where $\mu=1,2,3$ and $m (\bm p)\equiv m_0+m_2 p_\mu^2$.
Note that 
\begin{alignat}1
H^2 = A^2 p_\mu^2+m^2(\bm p)\equiv R^2(\bm p) \bm 1 .
\end{alignat}
Therefore, a deformed Hamiltonian 
\begin{alignat}1
\widetilde H\equiv \frac{H}{R}
\end{alignat}
has eigenvalues $\pm1$.
To characterise the topological property of $\widetilde H$,  
let us make a rotation in the $\tau$-space such that
$\tau_1\rightarrow\tau_2$, $\tau_2\rightarrow\tau_3$, $\tau_3\rightarrow\tau_1$.
Then, the Hamiltonian is converted into
\begin{alignat}1
\widetilde H&\rightarrow \frac{A \tau_2\sigma_\mu p_\mu+\tau_1 m (\bm p)}{R}
\nonumber\\
&=\frac{1}{R}
\left(\begin{array}{cc}&m (\bm p)-i A \sigma_\mu p_\mu \\ 
m (\bm p) + i A \sigma_\mu p_\mu& \end{array}\right)
\nonumber\\
&\equiv
\left(\begin{array}{cc}&Q (\bm p)\\ Q^\dagger (\bm p)& \end{array}\right).
\end{alignat}  
Here, a $2\times2$ SU(2) matrix $Q$ emerges, 
\begin{alignat}1
Q\equiv \frac{m (\bm p) - i A \sigma_\mu p_\mu}{R} ,
\end{alignat}
satisfying
$Q^\dagger Q=\bm 1$ and $\det Q=1$. Since $\pi_3(SU(2))=\mathbb{Z}$, 
$Q$ should be characterised by an integer winding number generically.

The winding number $\cal N$ is defined by
\begin{alignat}1
\cal N &=\frac{1}{24\pi^2}\int {\rm tr}\
\left(Q^\dagger d Q\right)^3
\nonumber\\
&=\frac{1}{24\pi^2}\int d^3 p\ 
\epsilon_{\mu\nu\rho}{\rm tr}\
(Q^\dagger \partial_\mu Q)(Q^\dagger \partial_\nu Q)(Q^\dagger \partial_\rho Q), 
\end{alignat}
where $d$ stands for the exterior derivative with respect to $p_\mu$, and $\partial_\mu$ is 
the derivative with respect to $p_\mu$.
For the normalization of the above equation, see, e.g., Eq. (66) of
Ref. \cite{Fukui_NPB}.

Let us look into the nature of this winding number more precisely.
Since $Q$ is a function of $p_\mu$, 
the winding number $\cal N$ characterizes the mapping from $\mathbb{R}^3$ to SU(2). 
However, at the infinity of $\mathbb{R}^3$, namely, $|p|\rightarrow\infty$, $Q$ becomes
a single element of SU(2), $Q\rightarrow {\rm sgn} (m_2)\bm 1$. Therefore, 
in the present mapping, the infinity $|p|\rightarrow \infty$ can be regarded as a
single point, which make it possible to regard $\mathbb{R}^3$ as $\mathbb{S}^3$. 
By the use of the fact that SU(2)$\sim \mathbb{S}^3$, the winding number $\cal N$ characterises
the mapping from $\mathbb{S}^3$ to $\mathbb{S}^3$, which can be classified by 
$\pi_3(\mathbb{S}^3)=\mathbb{Z}$. 

It is not difficult to guess the winding number in the following way. 
At the origin $|p|=0$, $Q={\rm sgn}(m_0)\bm 1$. Therefore, if ${\rm sgn}(m_0m_2)=1$,
$Q$ can be defomed into $Q={\rm sgn}(m_0)\bm 1$ by taking the limit $|m_0|\rightarrow\infty$, giving rise to a
trivial winding number ${\cal N} =0$. 
On the other hand, if ${\rm sgn}(m_0m_2)=-1$, $Q$ at $|p|=0$ cannot be deformed into $Q$ at $|p|=\infty$,
so that this case gives ${\cal N}=\pm1$.

Let us compute the winding number $\cal N$ concretely.
After a tedious but straightforward calucation, one finds
\begin{alignat}1
\epsilon_{\mu\nu\rho}
{\rm tr}\
&(Q^\dagger \partial_\mu Q)(Q^\dagger \partial_\nu Q)(Q^\dagger \partial_\rho Q)
\nonumber \\
&=12 \frac{\bm p^2 - m(\bm p)}{R^4}.
\end{alignat}
We reach, therefore, 
\begin{alignat}1
\cal N &=\frac{-12}{24\pi^2}\int_{{\rm R}^3} d^3 p \frac{m_0 - m_2 \bm p^2}{[\bm p^2+(m_0+m_2 \bm p^2)^2]^2}
\nonumber\\
&=\frac{-12\times (4\pi)}{24\pi^2}\int_0^\infty dp
\frac{p^2(m_0-m_2p^2)}{[p^2+(m_0+m_2 p^2)^2]^2}
\nonumber\\
&=-\frac{1}{2}\left[{\rm sgn}(m_0)-{\rm sgn}(m_2)\right].
\end{alignat}
The last formula indicates that the system is indeed 
in the topologically non-trivial (${\cal N} =\pm 1$) phase when $m_0$ and $m_2$ have
the {\it opposite} sign, 
albeit
in the trivial (${\cal N} =0$) phase when they have the {\it same} sign.

\section{The zero-energy condition}

The radial eigenvalue problem 
considered in Sec. II
has two basic ingredients: the eigenvalue
equation (\ref{radial}) and the boundary condition (\ref{bc}) at $r=R$
(on the surface of the sphere).
Here, we prove that 
the solutions of this radial boundary problem satisfies 
automatically the energy condition,
$E_\perp =0$.
This observation paves the way for constructing
the basis eigenspinors given in Eqs. (\ref{dv_basis}).
In other ``simpler'' geometries, 
such as the case a semi-infinite system with a flat boundary,
\cite{Shen_NJP}
or a cylindrical system infinitely long in the axial direction,
\cite{k2}
the scenario applies,
but the explicit use
of the zero-energy condition may not be
indispensable because of the simplicity of the problem.

We first need to find the explicit matrix form of
$H_\perp = H_{\rm bulk} |_{p_\theta=0, p_\phi=0}$.
Focus on the three momentum operators,
$p_\pm = -i \partial_\pm$,
$p_z = -i \partial_z$,
that have appeared in Eq. (\ref{matH}), and in addition,
$\bm p^2 = - \nabla^2$ in $m(\bm p)$.
In the spherical coordinates,
these can be expressed as
\begin{eqnarray}
\partial_\pm &=& 
e^{\pm i\phi} \left[ \sin \theta {\partial \over \partial r} 
+{1\over r}\left(
\cos \theta {\partial \over \partial \theta} 
\pm {i \over \sin \theta} {\partial \over \partial \phi} 
\right) \right]
\nonumber \\
&\equiv&
e^{\pm i\phi} \sin \theta {\partial \over \partial r} 
+ D_\pm,
\label{D_pm}
\end{eqnarray}
\begin{eqnarray}
\partial_z &=& \cos \theta {\partial \over \partial r} 
- {\sin \theta \over r}{\partial \over \partial \theta}
\nonumber \\
&\equiv& \cos \theta {\partial \over \partial r} + D_z,
\label{D_z}
\end{eqnarray}
and
\begin{equation}
\nabla^2 = {\partial \over \partial r}^2 + {2\over r}{\partial \over \partial r}
+{\bm L^2 \over r^2}, 
\label{nabla}
\end{equation}
where $\bm L^2$ is a square of the orbital angular momentum operator.
Since $D_\pm$, $D_z$ and $\bm L^2$ involve only angular derivatives,
we put them in $H_\parallel$.
Keeping only the first terms of Eqs. (\ref{D_pm}), (\ref{D_z}) and (\ref{nabla}), 
one finds,
\begin{equation}
H_\perp [\kappa] =
\left[
\begin{array}{cccc}
m_\perp [\kappa] & -i\kappa a  & 0 & -i\kappa b e^{-i\phi}  \\
-i\kappa a  & - m_\perp [\kappa] & -i\kappa b e^{-i\phi} & 0 \\
0 & -i\kappa b e^{i\phi} & m_\perp [\kappa] & i\kappa a \\ 
-i\kappa b e^{i\phi} & 0 & i\kappa a & - m_\perp [\kappa]   
\end{array}
\right],
\label{H_perp}
\end{equation}
where we have assumed a surface solution of the form of Eq. (\ref{damped}),
and replaced the $r$-derivatives with $\kappa$, the inverse of the penetration depth.
Eq. (\ref{H_perp}) can be indeed regarded as a $4\times 4$ $c$-number
matrix specified by a parameter $\kappa$.
We used the notation, $H_\perp [\kappa]$, to make this point explicit.
We have also introduced, for shortening the expression, the notations,
$a =A \cos\theta$, $b =A \sin\theta$ and
\begin{equation}
m_\perp [\kappa] = m_0 - m_2 \left(\kappa^2 + {2\over r}\kappa \right).
\label{m_perp}
\end{equation}
Here, in Eq. (\ref{m_perp}) an $r$-dependence in the last term looks cumbersome.
But as far as the surface wave function is well localized in the vicinity of the surface
at $r=R$, one can safely replace this coordinate $r$ by a constant $R$.
On the other hand, as far as the same assumption is applied
this last term itself becomes negligible,
since as far as $\kappa^{-1} \ll R$,
the second term is much larger than the last term.

Thanks to
the symmetric structure of the matrix form of Eq. (\ref{H_perp}),
the secular equation, 
$\det | H_\perp [\kappa] - E_\perp |=0$
for the radial eigenvalue problem (\ref{radial})
becomes as simple as
\begin{equation}
\det | H_\perp [\kappa] - E_\perp |
= \left\{ \kappa^2 A^2 - m_\perp [\kappa]^2 + E_\perp^2 \right\}^2 =0.
\end{equation}
This can be regarded as a quadratic equation for $\kappa^2$ 
under the approximation of
$m_\perp [\kappa] = m_0 - m_2 \kappa^2$,
with its two solutions $\kappa_\pm^2$ satisfying,
\begin{equation}
\kappa_+^2 \kappa_-^2 = {m_0^2 - E_\perp^2 \over m_2^2}.
\label{k+k-1}
\end{equation}
Now, in order to cope with the boundary condition (\ref{bc}),
two surface solutions of the form of Eq. (\ref{damped}),
one with $\kappa = \kappa_+$ and the other with $\kappa = \kappa_-$ 
must be superposed, i.e.,
\begin{equation}
|\psi \rangle = c_+ e^{\kappa_+ (r-R)} \bm u [\kappa_+] + c_- e^{\kappa_- (r-R)} \bm u [\kappa_-],
\label{psi_perp_A1}
\end{equation}
where
$\bm u [\kappa]$ is an eigenvector of $H_\perp [\kappa]$ given in Eq. (\ref{H_perp}).
The only way that this solution 
be compatible with the boundary condition (\ref{bc}) is
to have simultaneously
$c_+ +c_-=0$ and
$\bm u [\kappa_+] = \bm u [\kappa_-]$,
i.e.,
$|\psi \rangle$ takes the following form,
\begin{eqnarray}
|\psi \rangle &=& N \left[e^{\kappa_+ (r-R)} - e^{\kappa_- (r-R)} \right] \bm u [\kappa_+],
\nonumber \\
&\equiv& \rho (r) \bm u [\kappa_+],
\label{psi_perp_A2}
\end{eqnarray}
where $N$ is a normalization constant.

In the reduction from Eq. (\ref{psi_perp_A1})
to Eq. (\ref{psi_perp_A2}),
the second condition stating that two eigenvectors belonging to different $\kappa$'s
should coincide,
was crucial.
Indeed,
this coincidence occurs only under a very specific condition.
In order to clarify this point,
let us consider the following quantity,
\begin{eqnarray}
\Delta [\kappa] &\equiv& {H_\perp [\kappa] - E_\perp \over \kappa}
\nonumber \\
&=&\left[
\begin{array}{cccc}
M_1 [\kappa] & -i a  & 0 & -i b e^{-i\phi}  \\
-i a  & -M_2 [\kappa] & -i b e^{-i\phi} & 0 \\
0 & -i b e^{i\phi} & M_1 [\kappa] & i a \\ 
-i b e^{i\phi} & 0 & i a & -M_2 [\kappa]   
\end{array}
\right],
\label{Delta}
\end{eqnarray}
where
\begin{equation}
M_1 [\kappa] = {m_\perp [\kappa] - E_\perp \over \kappa},\ \ \
M_2 [\kappa] = {m_\perp [\kappa] + E_\perp \over \kappa}.
\end{equation}
Notice that
$\bm u [\kappa]$ is a zero-eigenvalue eigenvector of 
this matrix,
i.e.,
$\Delta [\kappa] \bm u [\kappa] = \bm 0$.
In order that
two of such eigenvectors $\bm u [\kappa]$
belonging to different $\kappa$'s ($=\kappa_\pm$), 
and therefore, to different $\Delta [\kappa]$'s
($\Delta [\kappa_+]$ and $\Delta [\kappa_-]$)
coincide,
both $M_1 [\kappa]$ and $M_2 [\kappa]$
must coincide.
Namely, one must have, simultaneously,
$M_1 [\kappa_+]=M_1 [\kappa_-]$ and
$M_2 [\kappa_+]=M_2 [\kappa_-]$.
Clearly,
a solution of the form of Eq. (\ref{psi_perp_A2}) is meaningful
only when $\kappa_+ \neq \kappa_-$.
Therefore,
$M_1 [\kappa_+]=M_1 [\kappa_-]$ signifies,
\begin{equation}
\kappa_+ \kappa_- = - {m_0 + E_\perp \over m_2},
\label{k+k-2}
\end{equation}
whereas
$M_2 [\kappa_+]=M_2 [\kappa_-]$ leads to
\begin{equation}
\kappa_+ \kappa_- = - {m_0 - E_\perp \over m_2}.
\label{k+k-3}
\end{equation}
Recalling that
Eqs. (\ref{k+k-2}) and (\ref{k+k-3}) must follow independently,
one can convince oneself that
the surface solution must satisfy the zero-energy condition,
\begin{equation}
E_\perp =0.
\label{zero}
\end{equation}
Note that Eqs. (\ref{k+k-2}) and (\ref{k+k-3}) are consistent with Eq. (\ref{k+k-1}),
but impose a stronger constraint on the values of $\kappa$'s and $E_\perp$.

\section{A brief reminder on the Jacobi's polynomials/differential equation}

\begin{itemize}

\item
The explicit form of the Jacobi's polynomials is given by the following
(differential) Rodrigues' formula:
\begin{eqnarray}
P_n^{\mu\nu} [\zeta] = {(-1)^n \over 2^n n!} {1\over \rho_{\mu\nu} [\zeta]}
{d^n \over dx^n} \left[ (1-\zeta^2)^n \rho_{\mu\nu} [\zeta] \right],
\label{Jac_R}
\end{eqnarray}
where
\begin{equation}
\rho_{\mu\nu} [\zeta] = (1-\zeta)^\mu (1-\zeta)^\nu.
\end{equation}
As is clear from its construction,
Eq. (\ref{Jac_R}) can be also expressed 
in the form of a contour integral
(the {\it integral} Rodrigues' formula).

\item 
The Jacobi's differential equation (\ref{Jac_D}) is a simple rewriting 
of the hypergeometric differential equation,
\begin{eqnarray}
\Big[
\xi (1-\xi) {d^2 \over d\xi^2}
+ \left\{ {\cal C} - (1+ {\cal A} + {\cal B}) \xi \right\} {d\over d\xi} &&
\nonumber \\
- {\cal A} {\cal B} 
\Big] F ({\cal A}, {\cal B}, {\cal C}, \xi)=0, &&
\label{hyper}
\end{eqnarray}
by the change of the independent variable,
\begin{equation}
\xi = {1-\zeta \over 2},
\end{equation}
and choice of the parameters,
\begin{eqnarray}
{\cal A} &=& -n,
\nonumber \\
{\cal B} &=& n+\mu +\nu +1,
\nonumber \\
{\cal C} &=& \mu +1.
\end{eqnarray}

\item
The Jacobi's polynomials
$P_n^{\mu\nu} [\zeta]$ 
satisfy the following ortho-normal relation,
\begin{eqnarray}
&& \int_{-1}^1 d\zeta
\rho_{\mu\nu} [\zeta]
P_{n_1}^{\mu\nu} [\zeta] P_{n_2}^{\mu\nu} [\zeta] 
\label{Jac_norm} \\
&=& \delta_{n_1 n_2}
{2^{\mu + \nu + 1} \Gamma (n_1 + \mu + 1) \Gamma (n_1 + \nu + 1)
\over n_1 ! (2n_1 + \mu + \nu + 1) \Gamma (n_1 + \mu + \nu + 1) }
\nonumber
\end{eqnarray}

\end{itemize}

\section{Proof of Eq. (\ref{cnm1})}
In order
to determine the relative magnitude and phase of
$c_{nm+}$ and $c_{nm-}$,
one needs to go back to Eqs. (\ref{couple})
and substitute $\beta_{m\sigma} [\zeta]$
given in Eqs. (\ref{beta2})
into this couple of equations
[naturally,
the change of the {\it dependent} variables
must be taken into account;
see Eq. (\ref{beta1})].
Changing the {\it independent} variable from $\theta$ to
$\zeta = \cos \theta$,
let us rewrite Eqs. (\ref{couple}) as
\begin{eqnarray}
\left[
\sqrt{1-\zeta^2}{d\over d\zeta} - {m+\zeta/2\over \sqrt{1-\zeta^2}}
\right]
\alpha_{m-} [\zeta]
&=& \lambda \alpha_{m+} [\zeta],
\nonumber \\
- \left[
\sqrt{1-\zeta^2}{d\over d\zeta} + {m-\zeta/2\over \sqrt{1-\zeta^2}}
\right]
\alpha_{m+} [\zeta]
&=& \lambda \alpha_{m-} [\zeta].
\end{eqnarray}
Performing the derivatives explicitly,
and changing the variables from $\alpha_{m\sigma}$
to $\beta_{m\sigma}$,
using Eq. (\ref{beta1}),
one finds
\begin{eqnarray}
\left[
(1-\zeta){d\over d\zeta} - \left(m+{1\over 2}\right)
\right]
\beta_{m-} [\zeta]
&=& \lambda \beta_{m+} [\zeta],
\nonumber \\
\left[
-(1+\zeta){d\over d\zeta} - \left(m+{1\over 2}\right) 
\right]
\beta_{m+} [\zeta]
&=& \lambda \beta_{m-} [\zeta]
\label{dbeta1}
\end{eqnarray}
for $m \ge 1/2$, and
\begin{eqnarray}
\left[
(1+\zeta){d\over d\zeta} - \left(m-{1\over 2}\right)
\right]
\beta_{m-} [\zeta]
&=& \lambda \beta_{m+} [\zeta],
\nonumber \\
\left[
-(1-\zeta){d\over d\zeta} - \left(m-{1\over 2}\right) 
\right]
\beta_{m+} [\zeta]
&=& \lambda \beta_{m-} [\zeta]
\label{dbeta2}
\end{eqnarray}
for $m \le 1/2$.
Recall that
$\beta_{m\sigma}$'s
as given in Eq. (\ref{beta2})
are proportional to the $n$-th order Jacobi's polynomial.
Eqs. (\ref{dbeta1}) and (\ref{dbeta2}) can be further simplified on account of
the following identities [{\it c.f.} Eqs. (A.7a) and (A.7b) of Ref. \cite{Abrikosov_arxiv}],
applicable to the derivative of the Jacobi's polynomials with a {\it specific}
choice of parameters $\mu$ and $\nu$
that are implied in these relations
through Eq. (\ref{beta2}), i.e.,
\begin{eqnarray}
&& (1-\zeta){d\over d\zeta} P_n^{m+{1\over 2}, m-{1\over 2}} =
\label{dp1} \\
&&
\left(m+{1\over 2}\right) P_n^{m+{1\over 2}, m-{1\over 2}}
- \left(n+m+{1\over 2}\right) P_n^{m-{1\over 2}, m+{1\over 2}},
\nonumber \\
&& - (1+\zeta){d\over d\zeta} P_n^{m-{1\over 2}, m+{1\over 2}} =
\label{dp2} \\
&&
\left(m+{1\over 2}\right) P_n^{m-{1\over 2}, m+{1\over 2}}
- \left(n+m+{1\over 2}\right) P_n^{m+{1\over 2}, m-{1\over 2}},
\nonumber
\end{eqnarray}
where $m \ge 1/2$.
These identities can be explicitly verified,
e.g., by the use of the integral counterpart
of Eq. (\ref{Jac_R}).

The final part of the proof of Eq. (\ref{cnm1})
lies in the comparison of 
Eqs. (\ref{dbeta1}), (\ref{dbeta2}) and
(\ref{dp1}), (\ref{dp2}).
For $m \ge 1/2$,
one can safely take off the operation of absolute value 
to the superscripts of Jacobi's polynomial
in Eq. (\ref{beta2}), yielding
\begin{equation}
\beta_{m\sigma} [\zeta] = \beta_{nm\sigma} [\zeta] 
= c_{nm\sigma} P_n^{m + {\sigma \over 2}, m - {\sigma \over 2}} [\zeta].
\end{equation}
Then, by simply
comparing Eqs. (\ref{dbeta1}) with (\ref{dp1}) and  (\ref{dp2}),
and recalling
$\lambda = \pm (n+m+1/2)$,
one can verify
$|c_{nm+}|=|c_{nm-}|$ with a relative sign of
\begin{equation}
c_{nm-}= - {\rm sign} [\lambda] c_{nm+}.
\label{cnm3}
\end{equation}
For $m \le -1/2$,
notice that
\begin{equation}
\left| m - {\sigma \over 2} \right| = |m| + {\sigma \over 2},
\end{equation}
i.e., for such $m$ Eq. (\ref{beta2}) becomes
\begin{equation}
\beta_{m\sigma} [\zeta] = \beta_{nm\sigma} [\zeta] 
= c_{nm\sigma} P_n^{|m| + {\sigma \over 2}, |m| - {\sigma \over 2}} [\zeta].
\end{equation}
This allows for the use of Eqs. (\ref{dp1}) and (\ref{dp2}) with $m$ replaced by $|m|$
in the couple of Eqs. (\ref{dbeta2}).
Taking note of $\lambda = \pm (n+|m|+1/2)$,
one can again verify $|c_{nm+}|=|c_{nm-}|$,
but this time with a relative sign of opposite value,
\begin{equation}
c_{nm-}= {\rm sign} [\lambda] c_{nm+}.
\label{cnm4}
\end{equation}
The relations (\ref{cnm3}) and (\ref{cnm4}), respectively,
for the two possible regimes of $m$
complete the proof of Eq. (\ref{cnm1}).

\bibliography{spherical_r4}

\end{document}